\documentclass[twocolumn,superscriptaddress,prl]{revtex4}
\usepackage[latin9]{inputenc}
\usepackage{amsmath}
\usepackage{amssymb}
\usepackage{graphicx}
\usepackage{wasysym}
\usepackage{esint}

\makeatletter
\@ifundefined{textcolor}{}
{%
 \definecolor{BLACK}{gray}{0}
 \definecolor{WHITE}{gray}{1}
 \definecolor{RED}{rgb}{1,0,0}
 \definecolor{GREEN}{rgb}{0,1,0}
 \definecolor{BLUE}{rgb}{0,0,1}
 \definecolor{CYAN}{cmyk}{1,0,0,0}
 \definecolor{MAGENTA}{cmyk}{0,1,0,0}
 \definecolor{YELLOW}{cmyk}{0,0,1,0}
}

\usepackage{color}

\makeatother

\begin{document}

\title{Observation and Control of Laser-Enabled Auger Decay}

\author{D. Iablonskyi}

\affiliation{Institute of Multidisciplinary Research for Advanced Materials, Tohoku
University, Sendai 980-8577, Japan}

\author{K. Ueda}

\email[]{ueda@tagen.tohoku.ac.jp}

\affiliation{Institute of Multidisciplinary Research for Advanced Materials, Tohoku
University, Sendai 980-8577, Japan}

\author{Kenichi L. Ishikawa}

\affiliation{Department of Nuclear Engineering and Management, Graduate School
of Engineering, The University of Tokyo, 7-3-1 Hongo, Bunkyo-ku, Tokyo
113-8656, Japan}

\affiliation{Photon Science Center, Graduate School of Engineering, The University
of Tokyo, 7-3-1 Hongo, Bunkyo-ku, Tokyo 113-8656, Japan}

\author{A. S. Kheifets}

\affiliation{Research School of Physics and Engineering, Australian National University,
Canberra, ACT 2601, Australia}

\author{P. Carpeggiani}

\affiliation{Dipartimento di Fisica, CNR-IFN, Politecnico di Milano, 20133 Milan,
Italy}

\author{M. Reduzzi}

\affiliation{Dipartimento di Fisica, CNR-IFN, Politecnico di Milano, 20133 Milan,
Italy}

\author{H. Ahmadi}

\affiliation{Dipartimento di Fisica, CNR-IFN, Politecnico di Milano, 20133 Milan,
Italy}

\author{A. Comby}

\affiliation{Dipartimento di Fisica, CNR-IFN, Politecnico di Milano, 20133 Milan,
Italy}

\author{G. Sansone}

\affiliation{Dipartimento di Fisica, CNR-IFN, Politecnico di Milano, 20133 Milan,
Italy}

\affiliation{Physikalisches Institut der Albert-Ludwigs-Universitat Stefan-Meier-Str.
19 79104 Freiburg, Germany}

\author{T. Csizmadia}

\affiliation{ELI-ALPS, Szeged, Hungary}

\author{S. Kuehn}

\affiliation{ELI-ALPS, Szeged, Hungary}

\author{E. Ovcharenko}

\affiliation{European XFEL, Hamburg, 22761 Germany}

\author{T. Mazza}

\affiliation{European XFEL, Hamburg, 22761 Germany}

\author{M. Meyer}

\affiliation{European XFEL, Hamburg, 22761 Germany}

\author{A. Fischer}

\affiliation{Max Planck Institute for Nuclear Physics, Heidelberg, 69117 Germany}

\author{C. Callegari}

\affiliation{Elettra-Sincrotrone Trieste, 34149 Basovizza, Trieste, Italy}

\author{O. Plekan}

\affiliation{Elettra-Sincrotrone Trieste, 34149 Basovizza, Trieste, Italy}

\author{P. Finetti}

\affiliation{Elettra-Sincrotrone Trieste, 34149 Basovizza, Trieste, Italy}

\author{E. Allaria}

\affiliation{Elettra-Sincrotrone Trieste, 34149 Basovizza, Trieste, Italy}

\author{E. Ferrari}

\affiliation{Elettra-Sincrotrone Trieste, 34149 Basovizza, Trieste, Italy}

\author{E. Roussel}

\affiliation{Elettra-Sincrotrone Trieste, 34149 Basovizza, Trieste, Italy}

\author{D. Gauthier}

\affiliation{Elettra-Sincrotrone Trieste, 34149 Basovizza, Trieste, Italy}

\author{L. Giannessi}

\affiliation{Elettra-Sincrotrone Trieste, 34149 Basovizza, Trieste, Italy}

\affiliation{ENEA C.R. Frascati, 00044 Frascati, Rome, Italy,}

\author{K. C. Prince}

\email[]{prince@elettra.trieste.it}

\affiliation{Elettra-Sincrotrone Trieste, 34149 Basovizza, Trieste, Italy}

\affiliation{Molecular Model Discovery Laboratory, Department of Chemistry and
Biotechnology, Swinburne University of Technology, Melbourne 3122,
Australia. }

\date{\today}
\begin{abstract}
Single photon laser enabled Auger decay (spLEAD) has been redicted
theoretically {[}Phys. Rev. Lett. 111, 083004 (2013){]} and here we
report its first experimental observation in neon. Using coherent,
bichromatic free-electron laser pulses, we have detected the process
and coherently controlled the angular distribution of the emitted
electrons by varying the phase difference between the two laser fields.
Since spLEAD is highly sensitive to electron correlation, this is
a promising method for probing both correlation and ultrafast hole
migration in more complex systems. 
\end{abstract}
\maketitle
\label{sec:1} When isolated atoms or molecules are excited, they
relax to their ground state in two ways: via nuclear motion or electronically,
and two possible ways of releasing electronic energy are by radiation
and by Auger decay. The Auger process \cite{1,2} first reported by
Lise Meitner, has played an important part in modern physics, particularly
surface science, because it is by far the strongest decay channel
for core holes of light elements such as carbon, nitrogen and oxygen.
However, for ionization of some inner valence levels of atoms and
molecules, the Auger process is energetically forbidden, because the
energy of the doubly ionized final state is higher than the energy
of the initial ion.

It was predicted some time ago that such states could decay by the
novel process of Interatomic/Intermolecular Coulombic Decay (ICD)
\cite{3}, if the excited or ionized species was weakly bound to its
environment. In this process, the final state of the system contains
two positive charges, but the charge is distributed between the original
atom or molecule and its neighbor, thus lowering the energy because
the Coulomb repulsion between the charges is reduced. This prediction
led subsequently to intense research activity, with many new discoveries
reported, as it was found that ICD and its variants could occur in
many different ways \cite{Jahnke2010,Mucke2010,Gokhberg2014,Trinter2014,Stumpf2016,You2016}.

Another recently reported exotic electronic decay process is Laser
Enabled Auger Decay (LEAD) \cite{Ranitovic2011}. Like ICD, LEAD is
a new decay mode of excited states, and both mechanisms depend on
the environment of the atom or molecule: in one case neighboring neutral
atoms or molecules, and in the second case, the surrounding photon
field. LEAD was observed for $3s$ ionized argon which normally decays
by fluorescence \cite{Ranitovic2011}. In the presence of a strong
infrared field, the ion can absorb multiple photons and decay to a
doubly $3p$ ionized state. Another class of LEAD, single photon LEAD
(spLEAD), has been theoretically predicted \cite{Cooper2013} but
so far has never been observed. Because spLEAD relies on electron
correlations, it can potentially provide novel information such as
correlation-induced ultrafast hole migration in molecules \cite{Cooper2013,Calegari2014,Kraus2015}.

In this work, we report the first observation of spLEAD, and moreover
we show that the emission can be coherently controlled. The process
itself can be viewed as a control of the decay of the ion by the laser
field. With coherent control, we also manipulate the angular distribution
of the emitted Auger electrons using the relative phase of the ionizing
and enabling laser beams.

The processes studied are illustrated graphically in Fig.~1. The
spLEAD takes place for the $2s$-hole state in Ne$^{+}$: 
\begin{equation}
2s^{1}2p^{6}+\omega\to2s^{2}2p^{4}(^{1}S,^{1}\!\!D)+{\rm e}^{-}\qquad.\label{eq:spLEAD}
\end{equation}
Within LS coupling, only the $^{1}S$ and $^{1}D$ final states are
allowed in spLEAD even though the $2s^{2}2p^{4}$ configuration can,
in principle, couple into the $^{3}P$ term. This is because the $2s^{1}2p^{6}$
is in the $S$ state and the Auger electron emitted from the $2p$
orbital is in the $s$ or $d$ state before the photoabsorption step
leading to spLEAD.

\begin{figure}[tb]
\includegraphics[width=1\columnwidth]{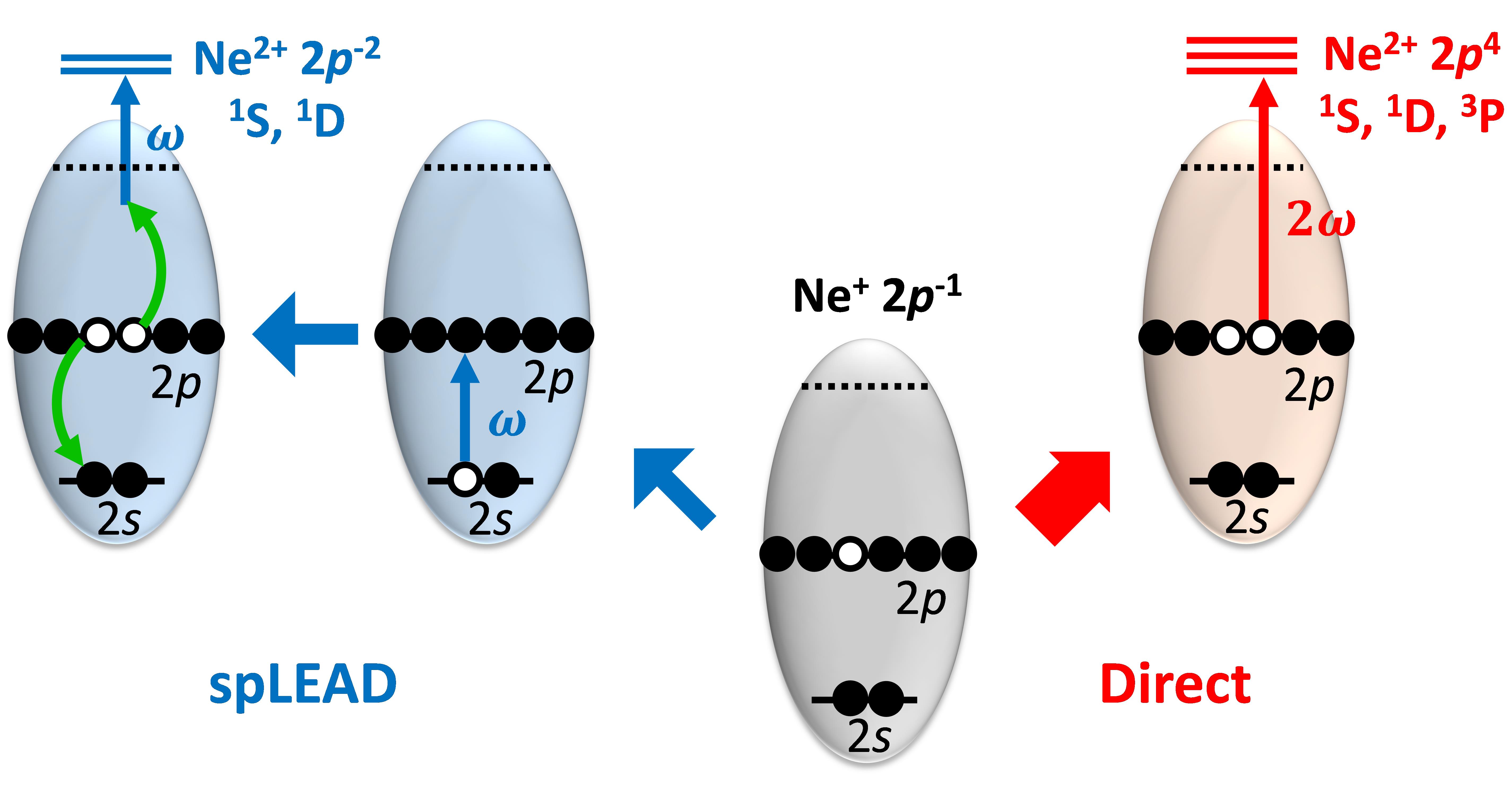} \caption{\label{fig:schematic diagram} Schematic diagram for ionization of
Ne $2s^{2}2p^{5}$. Center: initial state. Left (blue arrows): (1)
the spLEAD process follows (2) absorption of one $\omega$ photon;
Right (red arrows): direct ionization by $2\omega$ process (3).}
\end{figure}

The $2s$-hole state is prepared by absorption of a single photon
by the ionic ground state $2s^{2}2p^{5}\,{^{2}}P_{3/2}$ of Ne$^{+}$:
\begin{equation}
2s^{2}2p^{5}+\omega\to2s^{1}2p^{6},\label{eq:2s hole preparation}
\end{equation}
where $\omega=26.91$ eV is tuned to the energy difference between
$2s^{2}2p^{5}$ and $2s^{1}2p^{6}$. In addition, we simultaneously
ionize the ground state $2s^{2}2p^{5}$ of Ne$^{+}$ by the phase-locked
second harmonic $2\omega$ (53.82 eV) to access all of the $^{1}S$,
$^{1}D$ and $^{3}P$ final states, Ne$^{2+}$ \quad{}%
\mbox{%
(Fig.~\ref{fig:schematic diagram}, right)%
}: 
\begin{equation}
2s^{2}2p^{5}+2\omega\to2s^{2}2p^{4}(^{1}S,^{1}\!\!D,^{3}\!\!P)+{\rm e}^{-}.\label{eq:direct ionization}
\end{equation}
The $2s^{2}2p^{5}$ ionic states can be directly ionized by process
(\ref{eq:direct ionization}), or indirectly by processes (\ref{eq:2s hole preparation})
and (\ref{eq:spLEAD}) \quad{}%
\mbox{%
(Fig.~\ref{fig:schematic diagram}, left)%
}: 
\begin{align}
2s^{2}2p^{5}+\omega+\omega & \to2s^{1}2p^{6}+\omega\label{eq:spLEAD path}\\
 & \to2s^{2}2p^{4}(^{1}S,^{1}\!\!D)+{\rm e}^{-}.\nonumber 
\end{align}
The key point is that the two light pulses $\omega$ and $2\omega$
are phase coherent. As a result, the two paths (\ref{eq:direct ionization})
and (\ref{eq:spLEAD path}) to the same final states $^{1}S$ and
$^{1}D$ interfere, but only path (\ref{eq:direct ionization}) leads
to $^{3}P$, so no interference occurs.

To investigate the spLEAD process exhibited in Fig.~\ref{fig:schematic diagram},
we chose to use the recently demonstrated capability of the FERMI
light source to produce intense bichromatic radiation with controllable
phase \cite{Prince2016}. We provided conditions in which there are
two quantum paths to a photoelectron state with a defined linear momentum,
and we coherently controlled these quantum paths. With this approach,
we observed a signal proportional to the amplitude, rather than intensity,
of the quantum processes. This approach brings considerable advantages
when detecting a weak signal with a strong background \cite{Gunawardena_Elliott}.

We illustrate here how the spLEAD emission can be coherently controlled
and detected in the photoelectron angular distribution (PAD), using
that of the $^{1}S$ final state as an example. Process (4) emits
a $p$ wave electron while process (3) is assumed to emit mainly a
$d$ wave electron, by the Fano propensity rule \cite{PhysRevA.32.617}.
The angular distribution can then be written as: 
\begin{equation}
I(\theta,\phi)\propto\left|Y_{20}(\theta,\phi)e^{i\eta}+\sqrt{c}\,Y_{10}(\theta,\phi)e^{i\delta_{pd}}\right|^{2}\ ,\label{eq:PAD}
\end{equation}
where $Y_{nl}(\theta,\phi)$ denote spherical harmonics, and $c(\ll1)$
is proportional to the intensity of the spLEAD path (\ref{eq:spLEAD path})
relative to the direct photoionization (\ref{eq:direct ionization}),
$\delta_{pd}$ is their phase difference, and $\eta$ is the $\omega-2\omega$
relative phase. Note that the relative intensity enters the equation
as the term $\sqrt{c}$, that is, as the relative amplitudes of the
two coherent beams. Then the asymmetry of the electron emission, defined
as the difference between the emission in one hemisphere ($0<\theta<\pi/2$)
and the other ($\pi/2<\theta<\pi$), divided by the sum, is expressed
as 
\begin{equation}
A(\eta)=\frac{\sqrt{15}}{4}\frac{\sqrt{c}}{1+c}\cos(\eta-\delta_{pd}).\label{eq:asymmetry}
\end{equation}
This oscillates as a function of the relative phase $\eta$ between
the two harmonics, and its amplitude is approximately $\sqrt{c}(\gg c)$
instead of $c$.

The experiment was performed at the Low Density Matter beamline \cite{Lyamaev}
of FERMI. The sample consisted of a mixture of neon and helium (for
calibration purposes) and was exposed to a bichromatic beam of temporally
overlapping first and second harmonic radiation with controlled phase
relationship \cite{Prince2016}. In the present experiment, the FEL
fundamental wavelength was generated by tuning the sixth undulator
of the radiator to the 5th harmonic of the seed wavelength (230.4
nm). The second harmonic of the FEL was generated by tuning the first
five undulators to the 10th harmonic of the seed, giving rise to bichromatic
phase-locked pulses. The photon energies were $\omega=26.91$ eV and
$2\omega=53.82$ eV; where $\omega$, as stated, was equal to the
difference in energy between the $2s^{2}2p^{5}$ and $2s^{1}2p^{6}$
Ne$^{+}$ ionic states (see Fig.~\ref{fig:schematic diagram}). The
light was focused to a spot size of approximately 5 $\mu$m FWHM.
The calculated pulse durations were 75 fs for the first and 60 fs
for the second harmonic.

Under the above conditions, the kinetic energies of the photoelectrons
emitted by the fundamental $\omega=26.91$ eV (from the $2p$ sub-shell)
and by the second harmonic $2\omega=53.82$ eV (from the $2s$ sub-shell)
are identical. Furthermore, similar ionization rates of $2p$ by the
fundamental ($\omega$) and 2s by the second harmonic ($2\omega$)
were set. Single ionization generates our sample, which is a mixture
of $2s^{2}2p^{5}$ and $2s^{1}2p^{6}$ Ne$^{+}$ ions formed by three
different processes: $2p$ ionization by $\omega$: 
\begin{equation}
2s^{2}2p^{6}+\omega\to2s^{2}2p^{5}+{\rm e}^{-},\label{eq:2p ionization by w}
\end{equation}
$2p$ ionization by $2\omega$: 
\begin{equation}
2s^{2}2p^{6}+2\omega\to2s^{2}2p^{5}+{\rm e}^{-},\label{eq:2p ionization by 2w}
\end{equation}
and $2s$ ionization by $2\omega$: 
\begin{equation}
2s^{2}2p^{6}+2\omega\to2s^{1}2p^{6}+{\rm e}^{-}.\label{eq:2s ionization by 2w}
\end{equation}
These ionic states, which are our target initial states of processes
(1)-(4), are independent and have no quantum phase relationship.

The Ne-He mixture was introduced into the experimental chamber using
a pulsed valve (Parker model 9, convergent-to-cylindrical nozzle with
flat aperture of 250 micron diameter) at room temperature. Helium
was added as a calibrant and as a cross-check for spurious artefacts.
The electron spectra were measured using a velocity map imaging (VMI)
spectrometer, and the spectra reconstructed from the raw data using
the pBASEX algorithm \cite{Garcia2004}.

The spectrum (at fixed relative phase) is shown in Fig. 2, together
with the line assignments. Several features are present in the spectrum,
of which the most prominent are the single-photon emission of $2p$
and $2s$ electrons at 5.3 eV (by $\omega$ and $2\omega$ respectively;
the photoelectrons have the same kinetic energy), single photon ionization
of $2p$ by $2\omega$ (32.2 eV), and emission corresponding to doubly
ionized Ne$^{2+}$ final states $^{1}S$, $^{1}D$ and $^{3}P$.

\begin{figure}[tb]
\includegraphics[width=1\columnwidth]{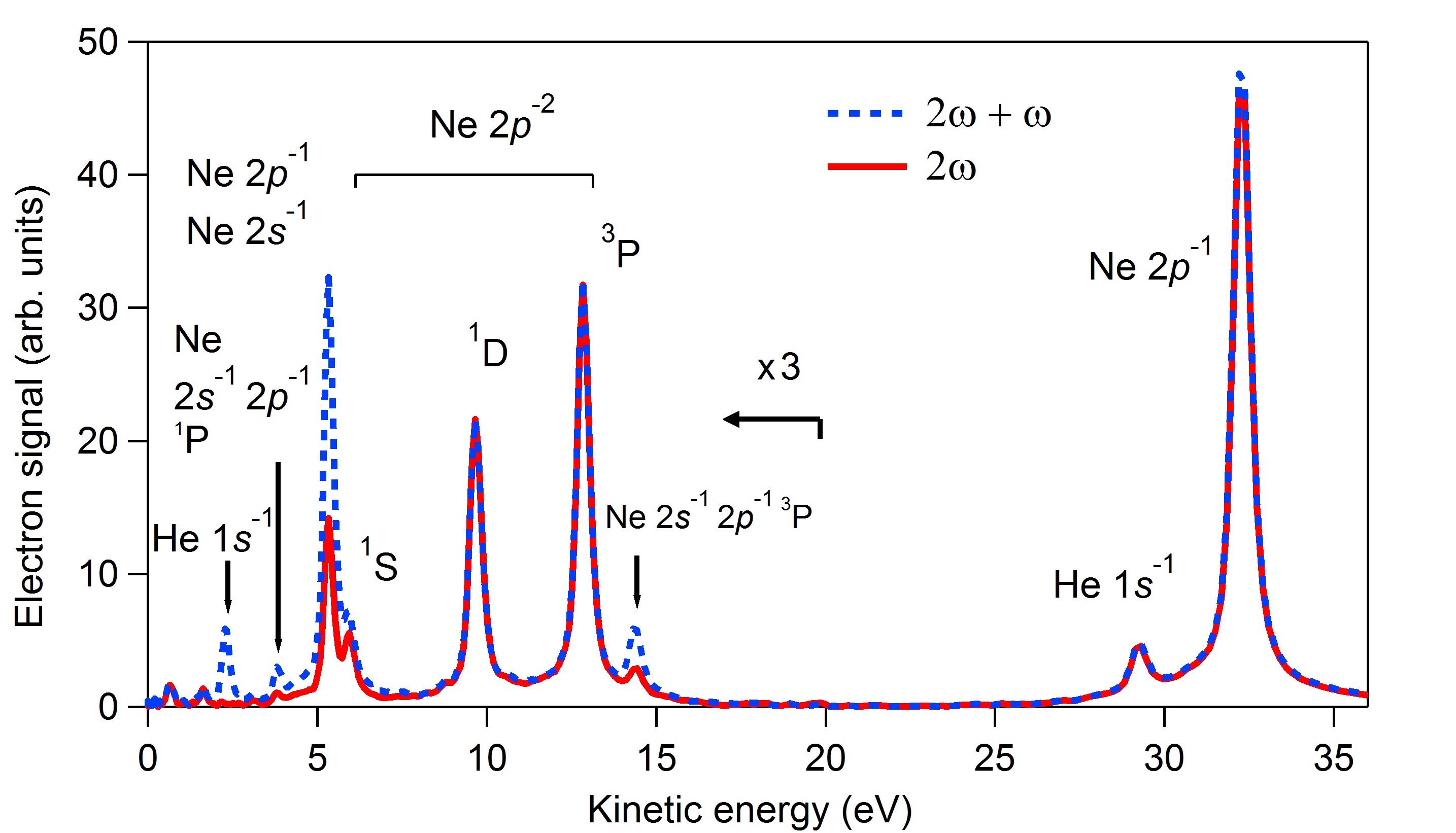} \caption{\label{fig:spectra} (Color online.) Electron kinetic energy spectra
of atomic Ne irradiated by $\omega=26.91$ eV and $2\omega=53.82$
eV (dashed blue curve) and $2\omega$ only (continuous red curve).}
\end{figure}

In principle, spLEAD may be observed directly as an increase in the
photoelectron yield of the $^{1}S$ and $^{1}D$ final states when
the $\omega$ field is applied, Eq. 4. However, the increase ($\sim0.05\%$
in the present case from the estimation described below) is too small.
Indeed, within experimental error, the intensities of the $^{1}$S
and $^{1}$D peaks in Fig. 2 did not change when the first harmonic
was added. In the PAD, however, we observed an oscillation of the
asymmetry $A(\eta)$ as a function of phase difference between the
harmonics, see Fig.~3. Note that the zero of the relative phase is
not absolute, but has an arbitrary offset. We observe strong modulation
of the $2p$$^{4}$ $^{1}S$ and $^{1}D$ states as a function of
phase, while the $^{3}P$ state shows much weaker or negligible modulation.
The helium and neon single ionization peaks show no effect, as expected.
The oscillation in the PADs indicates that indeed the final states
can be reached from an initial state by more than one pathway, and
that these pathways are coherent.

\begin{figure}[tb]
\centering \includegraphics[width=0.7\columnwidth]{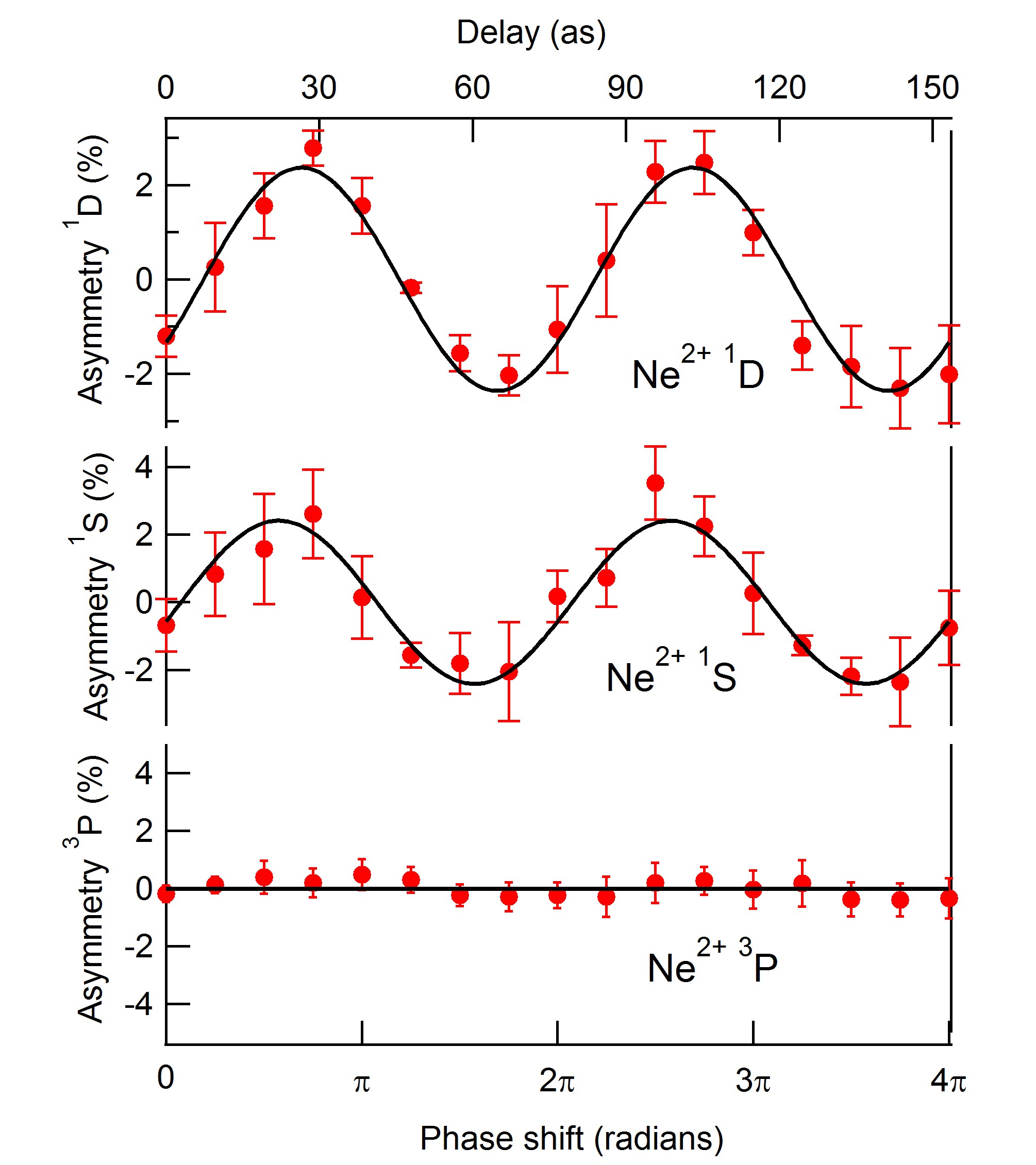} \caption{\label{fig:asymmetry} Asymmetry $A(\eta)$ of photoelectron angular
distributions corresponding to final ionic states$^{1}S$, $^{1}D$
and $^{3}P$ of Ne$^{2+}$, as a function of phase shift between first
and second harmonics. The asymmetry $A(\eta)$ is defined as the ratio
between the difference and the sum of the emission in one hemisphere
($0<\theta<\pi/2$) and the other ($\pi/2<\theta<\pi$).}
\end{figure}

We note that our measurement belongs to the Brumer-Shapiro class of
bichromatic experiments. In such experiments, the phase-locked first
and second harmonics create two quantum paths for the emission of
a photoelectron of given momentum, causing interference \cite{Brumer_Shapiro},
for example, by single-photon and two-photon ionization of the same
electron state. The conditions we have chosen also set up interference,
but in a quite different way from the usual Brumer-Shapiro approach.
Here the interference appears as a phase dependent asymmetry of the
PADs as given by Eq.~(\ref{eq:asymmetry}).

In order to confirm that the observed oscillations stem from coherently
controlled spLEAD, we estimate the amplitudes for the oscillation
of $A(\eta)$ in Fig.~3. We use the experimental conditions captured
in the photoelectron spectra in Fig.~\ref{fig:spectra}, and the
calculated spLEAD cross section. From the spectrum in which only the
second harmonic was present (dashed blue curve in Fig.~\ref{fig:spectra}),
we extract the intensities of the photoelectron peaks corresponding
to processes (\ref{eq:direct ionization}), (8) and (9), as well as
the following process: 
\begin{equation}
2s^{1}2p^{6}+2\omega\to2s^{1}2p^{5}(^{1,3}P)+{\rm e}^{-}\label{eq:2s1 2p5}
\end{equation}
with electron kinetic energies of 3.85 eV (${^{1}}P$) and 14.4 eV
(${^{3}}P$). In ion spectra (data not shown), the ratio of ${\rm Ne}^{+}/{\rm Ne}^{2+}$
was found to be 3.85. This was combined with the known photoionization
cross sections {[}6.7, 6.8, and 0.3 Mb for (3), (8), and (9), respectively,
and estimated cross section of 6.8 Mb for (10){]}, to give an estimate
of the intensity of $2\omega$ of $2.0\times10^{13}$ W/cm$^{2}$.
This was done by solving the rate equations and taking account of
the pulse duration and the spot size given above; Gaussian temporal
and spatial profiles were used. From the spectrum (continuous red
curve in Fig.~\ref{fig:spectra}), where fundamental radiation $\omega$
was added without altering the intensity of $2\omega$, we again extract
photoelectron intensities including that for process (7). From the
intensity ratios of photoelectrons of processes (7)-(9) and known
cross sections for these processes {[}8.4 Mb for (7){]}, we estimate
the intensity ratio between $\omega$ and $2\omega$ to be 1:52, and
hence, the intensity of $\omega$ to be $3.8\times10^{11}$ W/cm$^{2}$.

Furthermore, as can be seen in Fig.~\ref{fig:spectra}, the addition
of $\omega$ leads to a significant enhancement of the overlapping
peak of $2p$ and $2s$ photoelectrons for the processes (7) and (9)
respectively, and of the intensity for the process (10). These observations
clearly illustrate that the fundamental radiation $\omega$ indeed
enhances the population of $2s^{1}2p^{6}$ via the $2s$-$2p$ hole
coupling Eq.~(\ref{eq:2s hole preparation}). From the magnitude
of the enhancement of the photoelectron peak arising from the process
(\ref{eq:2s1 2p5}), relative to the photoelectron peak intensities
of the processes (7) and (8), we estimate the effective probability
of process (\ref{eq:2s hole preparation}) to be 0.065.

The amplitude for the oscillation of $A(\eta)$ in Fq. (6) is equal
to the amplitude of the spLEAD process (4) relative to the direct
ionization process (2), i.e. $\HF\sqrt{c}$ in Eq. (5); both follow
practically the same first step ionization dominated by process (8).
The only missing parameter needed to estimate $\sqrt{c}$ is therefore
the amplitude of the spLEAD process (1). The estimate of $\sqrt{c}$
employing the spLEAD cross section reported in the literature \cite{Cooper2013},
however, resulted in $\sqrt{c}\HF0.001$, roughly one order smaller
than the present observation $\sqrt{c}\HF0.02$. Our own calculations
including only the static correlations (i.e. configuration mixing
in $2s$$^{1}$$2sp$$^{6}$ $^{2}$$S$$_{1/2}$) are in good agreement
with the literature value. We thus extended the calculations to include
the dynamic correlations (continuum-continuum coupling) using the
the following expression:

\begin{equation}
M(\nu_{f},\nu_{0};l_{n})=\left\{ \sum_{\varepsilon_{\nu_{n}}<0}+\int_{0}^{\infty}\!\!\!d\varepsilon_{\nu_{n}}\right\} \frac{d_{\nu_{f}\nu_{n}}(\omega)\,D_{\nu_{n}\nu_{0}}}{\omega_{n0}-\omega-i\delta}\ .
\end{equation}
Here $\nu_{0}$ is the initial $2s$ state, $\nu_{n}$ is an intermediate
state $2p^{4}n\ell$ or $2p^{4}\varepsilon\ell$, $\ell=0,2$ and
$\nu_{f}$ is the $\epsilon p$ or $f$ final state, $\omega_{n0}$
is the energy difference between the intermediate and initial state.
The Auger decay matrix element $D_{\nu_{n}\nu_{0}}(\omega_{j})$ is
calculated using the computer code described in \cite{Amusia1997}.
The discrete sum in this equation is equivalent to inclusion of the
static correlation in \cite{Cooper2013}. The dipole matrix elements
$d_{\nu_{f}\nu_{n}}$ for the continuum-continuum transitions are
singular. Their integration is handled using the numerical recipe
\cite{Korol1993}

The oscillation amplitudes calculated using these estimates are 0.82\%
for the $^{1}S$ peak and 1.4\% for $^{1}D$, in rough agreement with
our observation, 2.40 $\pm${} 0.34\% for $^{1}S$ and 2.33 $\pm${}
0.18\% for $^{1}D$. The discrepancy may stem from the experimental
uncertainties and theoretical approximations involved. Note that here,
unlike in Eqs.~(\ref{eq:PAD}) and (\ref{eq:asymmetry}), more exact
treatments, with inclusion of the weaker $s$ wave for process (\ref{eq:direct ionization})
and taking account of the $^{1}D$ final state, are used. The observed
oscillations of the $^{1}S$ and $^{1}D$ asymmetry show a phase difference
of 0.36$\pm$ 0.14 rad, in qualitative agreement with the predictions
of theory, but significantly smmaller than the calculated value of
1.27 rad. The discrepancy is most likely due to the limits of the
accuracy with which relative phases and matrix elements can be calculated.

In summary, we have shown how the technique of coherent control can
be applied to observe a new phenomenon, spLEAD, which for the present
case of neon is too weak to be observed directly as an intensity enhancement.
We detected the process by using a method which depends on amplitude
rather than intensity. Furthermore, we manipulated the outcome of
the ionization process, in terms of the emission direction of the
photoelectrons. Cooper and Averbukh \cite{Cooper2013} calculated
that oxygen $2s$ holes of glycine have a far higher cross-section
for spLEAD, implying detection is much easier. Our results suggest
that spLEAD may be observable as an increase in cross-section for
molecules containing oxygen, which includes many biomolecules. Noting
that sudden creation of the O $2s$ hole causes ultrafast charge dynamics,
our work shows the way to investigating charge dynamics not only by
the pump-probe methods proposed by Cooper and Averbukh \cite{Cooper2013},
but also by the sophisticated methods of coherent control. The complex
and ultrafast dynamics of inner-valence hole wavepackets, or electron
correlations, may now be investigated and coherently controlled by
using the coherence of bichromatic light with resolution of a few
attoseconds.

\section*{Acknowledgments}

This work was supported in part by the X-ray Free Electron Laser Utilization
Research Project and the X-ray Free Electron Laser Priority Strategy
Program of the Ministry of Education, Culture, Sports, Science, and
Technology of Japan (MEXT) and IMRAM program of Tohoku University,
and Dynamic Alliance for Open Innovation Bridging Human, Environment
and Materials program. K.L.I. gratefully acknowledges support by the
Cooperative Research Program of ``Network Joint Research Center for
Materials and Devices (Japan)'', Grant-in-Aid for Scientific Research
(No. 25286064, 26600111, and 16H03881) from MEXT, the Photon Frontier
Network Program of MEXT, the Center of Innovation Program from the
Japan Science and Technology Agency, JST, and CREST (Grant No.~JPMJCR15N1),
JST. We acknowledge the support of the Alexander von Humboldt Foundation
(Project Tirinto), the Italian Ministry of Research (Project FIRB
No. RBID08CRXK and PRIN 2010ERFKXL\_006), the bilateral project CNR-JSPS
``Ultrafast science with extreme ultraviolet Free Electron Lasers'',
and funding from the European Union Horizon 2020 research and innovation
programme under the Marie Sklodowska-Curie grant agreement No. 641789
MEDEA (Molecular Electron Dynamics investigated by IntensE Fields
and Attosecond Pulses). We thank the machine physicists of FERMI for
making this experiment possible by their excellent work in providing
high quality FEL light.

\end{document}